\documentclass[%
reprint,
superscriptaddress,
showkeys,
amsmath,
amssymb,
aps,
prb,
floatfix,
]{revtex4-2}

\usepackage[none]{hyphenat}
\usepackage{graphicx}% Include figure files
\usepackage{dcolumn}% Align table columns on decimal point
\usepackage{bm}% bold math
\usepackage{hyperref}
\usepackage{siunitx}
\hypersetup{
	colorlinks=true,
	linkcolor=black,
	filecolor=black,      
	urlcolor=blue,
	citecolor=black,
	}
\usepackage{url}
\usepackage[nameinlink]{cleveref}
\usepackage[version=4]{mhchem}
\usepackage{color}
\usepackage{changes}

\begin{document}

\preprint{APS/123-QED}

\title{The role of Debye temperature in achieving large adiabatic temperature changes at cryogenic temperatures: a case study on \ce{Pr2In}}

\author{Wei Liu}
\email{wei.liu@tu-darmstadt.de}
\author{Franziska Scheibel}
\author{Nuno Fortunato}
\author{Imants Dirba}
\affiliation{Institute of Materials Science, Technical University of Darmstadt, 64287 Darmstadt, Germany }
\author{Tino Gottschall}
\affiliation{Dresden High Magnetic Field Laboratory (HLD-EMFL), Helmholtz-Zentrum Dresden-Rossendorf, 01328 Dresden, Germany}
\author{Hongbin Zhang}
\author{Konstantin Skokov}
\author{Oliver Gutfleisch}

\affiliation{Institute of Materials Science, Technical University of Darmstadt, 64287 Darmstadt, Germany }
\date{\today}

\begin{abstract}
The excellent magnetic entropy change ($\Delta S_T$) in the temperature range of 20 $\sim$ \qty{77}{\kelvin} due to the first-order phase transition makes \ce{Pr2In} an intriguing candidate for magnetocaloric hydrogen liquefaction. As an equally important magnetocaloric parameter, the adiabatic temperature change ($\Delta T_{ad}$) of \ce{Pr2In} associated with the first-order phase transition has not yet been reported. In this work, the $\Delta T_{ad}$ of \ce{Pr2In} is obtained from heat capacity measurements: 2 K in fields of 2 T and \qty{4.3}{\kelvin} in fields of 5 T. While demonstrating a $\Delta T_{ad}$ that is not as impressive as its remarkable $\Delta S_T$, \ce{Pr2In} exhibits a low Debye temperature ($T_D$) of around 110 K. Based on these two observations, an approach that combines the mean-field and Debye models is developed to study the correlation between $\Delta T_{ad}$, one of the most important magnetocaloric parameters, and $T_D$, one important property of a material. The role of $T_D$ in achieving large $\Delta T_{ad}$ is revealed: materials with higher $T_D$ tend to exhibit larger $\Delta T_{ad}$, particularly in the cryogenic temperature range. This discovery explains the absence of an outstanding $\Delta T_{ad}$ in \ce{Pr2In} and can serve as a tool for designing or searching materials with both a large $\Delta S_T$ and a $\Delta T_{ad}$. 
\end{abstract}

\keywords{Magnetism, Magnetic Materials, Phase transition, Magnetocaloric effect, Thermodynamics}

\maketitle

\section{\label{sec:level1}Introduction}

Magnetocaloric materials with large isothermal magnetic and adiabatic temperature changes ($\Delta S_T$ and $\Delta T_{ad}$) in the temperature range from 20 K (condensation point of \ce{H2}) to 77 K (condensation point of \ce{N2}) are required for the successful implementation of magnetocaloric hydrogen liquefaction \cite{Liu.2023, Liu.2022, Tang.2022, Zhang.2019, Koshkidko.2023}, an emerging cooling technology based on the magnetocaloric effect with great potential to achieve higher efficiency than the conventional liquefaction methods based on Joule-Thomson expansion \cite{Kitanovski.2020, Matsumoto.2011, Feng.2020, Barclay.2019}. In this sense, rare-earth-based intermetallic compounds are promising candidates for magnetocaloric hydrogen liquefaction \cite{Liu.2023,Liu.2022,Franco.2018,Zheng.2017,Law.2023,Li.2020}. In particular, the heavy rare-earth-based (Gd, Tb, Dy, Ho, Er, and Tm) ones such as \ce{HoB2} \cite{Castro.2020}, \ce{ErAl2} \cite{Yang.2023}, and \ce{HoAl2} \cite{Gil.2016} have been intensively investigated due to their large magnetocaloric effects within the temperature range of 20 $\sim$ \qty{77}{\kelvin}. \par

Although light rare-earth elements (La, Ce, Pr, Nd, and Sm) typically have a much lower resource criticality than heavy rare-earth elements and therefore are more suitable for large-scale applications of magnetocaloric hydrogen liquefaction, light rare-earth-based intermetallic compounds are often overlooked because they generally show a weaker magnetocaloric effect than their heavy rare-earth counterparts \cite{Liu.2023}. The larger magnetocaloric effects of heavy rare-earth-based materials are attributed to the larger magnetic moments of heavy rare-earth ions \cite{Liu.2023}. The light rare-earth ions, namely \ce{Ce^3+}, \ce{Pr^3+}, \ce{Nd^3+}, and \ce{Sm^3+}, have a magnetic moment below 4 $\mu_B$, much smaller than the heavy rare-earth ions of \ce{Gd^3+}, \ce{Tb^3+}, \ce{Dy^3+}, \ce{Ho^3+}, \ce{Er^3+}, and \ce{Tm^3+}, which show a magnetic moment greater than \qty{7}{\mu_B} \cite{Coey.2009}. \par

However, the report on \ce{Pr2In} showing an excellent $\Delta S_T$ of about \qty{20}{\joule\per\kelvin\per\kilogram} in magnetic fields of \qty{5}{\tesla} at about \qty{57}{\kelvin} \cite{Biswas.2022} opens a new pathway that breaks the aforementioned “stereotype”. Although known for demonstrating the strongest magnetocaloric effect among the heavy rare-earth \ce{R2In} (R: Gd, Tb, Dy, Ho, and Er) system \cite{Zhang.2011}, the second-order magnetocaloric material \ce{Er2In} with a Curie temperature ($T_C$) of \qty{20}{\kelvin} exhibits a $\Delta S_T$ of \qty{15.5}{\joule\per\kelvin\per\kilogram}, significantly smaller than \ce{Pr2In}. The giant $\Delta S_T$ within the temperature range of 20 $\sim$ 77 K makes \ce{Pr2In} an attractive candidate for magnetocaloric hydrogen liquefaction. \par

The giant $\Delta S_T$ in \ce{Pr2In} is ascribed to its first-order magnetic phase transition \cite{Biswas.2022,Biswas.2020}. This alloy, as well as \ce{Nd2In} and \ce{Eu2In}, was initially reported to show a first-order phase transition by Forker \textit{et al.} in 2005, evidenced by the measurements of magnetic and electric hyperfine interactions \cite{Forker.2005}. Subsequently, in 2018 Guillou \textit{et al.} reported the giant first-order magnetocaloric effect in \ce{Eu2In} \cite{Guillou.2018}, triggering a series of experimental and theoretical studies on this compound and its relatives \cite{Ryan.2019,MendiveTapia.2020,Guillou.2020,Alho.2020}. It is worth mentioning that Tapia-Mendive \textit{ et al.} theoretically demonstrated that the first-order phase transition in \ce{Eu2In} is due to a topological change to the Fermi surface. \cite{MendiveTapia.2020}. \par

Soon after the observation of the giant $\Delta S_T$ in \ce{Eu2In}, the excellent $\Delta S_T$ in \ce{Pr2In} \cite{Biswas.2020} and \ce{Nd2In} \cite{Liu.2021,Biswas.2022b} were reported. It is worth mentioning that there is also a study reporting that \ce{Pr2In} exhibits a second-order phase transition without a significantly large $\Delta S_T$ \cite{Cui.2022}. The reason for this discrepancy is not yet clear and could be attributed to differences in sample preparation and heat treatment. \par

Despite the fact that $\Delta T_{ad}$ is as important as $\Delta S_T$ for the magnetocaloric effect \cite{Gottschall.2019b}, $\Delta T_{ad}$ of \ce{Pr2In} showing a first-order magnetic phase transition remains unreported. The first part of our work is about revisiting \ce{Pr2In} to obtain its $\Delta T_{ad}$ by constructing the total entropy curves from heat capacity data. The discoveries of the absence of an outstanding $\Delta T_{ad}$ and the low Debye temperature ($T_D$) in \ce{Pr2In} motivate us to study the correlation between $\Delta T_{ad}$ and $T_D$ to explain why \ce{Pr2In} shows no remarkable $\Delta T_{ad}$ and explore ways to improve this important magnetocaloric parameter. \par

\section{Experiment}
\ce{Pr2In} was synthesized by arc-melting high-purity raw materials Pr (99.5 wt.\% pure) and In (99.99 wt.\% pure) five times. To ensure good homogeneity, the ingot was flipped after each melting. As the surface of the \ce{Pr2In} sample reacts with air, the ground powder was sealed in a capillary hermetically in an Ar-filled glovebox (p(O$_2$)$<$0.1 ppm) for powder X-ray diffraction (XRD). The powder XRD measurement was performed using a powder diffractometer (Stadi P, Stoe \& Cie GmbH) equipped with a Ge111-Monochromator using MoK$\alpha _1 $-radiation ($\lambda$ = 0.70930 \r{A}) in the Debye–Scherrer geometry. Magnetization as a function of temperature in magnetic fields up to 10 T were measured by a Physical Properties Measurement System (PPMS) from Quantum Design. Heat capacity in magnetic fields of 0, 1, 2, 5 and 10 T was measured in the same PPMS with the 2$\tau$ approach. \par

\section{Results and discussion}

\subsection{Phase purity}

The sufficient purity of the \ce{Pr2In} crystallizing in \ce{Ni2In}-type hexagonal structure (space group: P6\textsubscript{3}/mmc) is confirmed by the XRD measurement. The XRD patterns and the results of Rietveld refinement are included in the supplementary \cite{Supplementary}.

\subsection{Magnetocaloric properties}
\begin{figure*}[ht!]
    \centering
    \includegraphics[trim={0 2cm 0 2cm},clip, width = \linewidth]{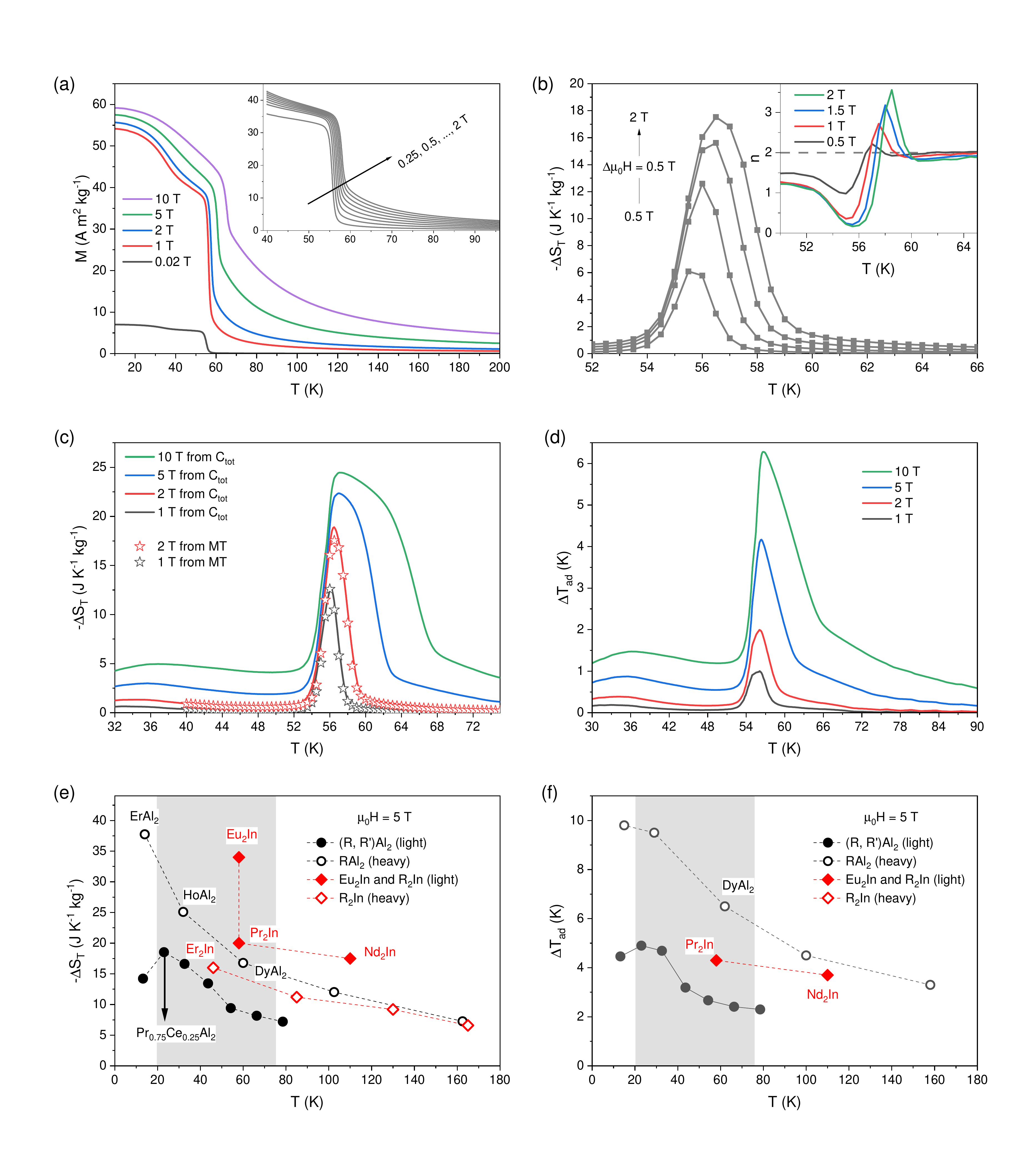}
    \caption{ (a) Magnetization of \ce{Pr2In} as a function of temperature. (b) $\Delta S_T$ of Pr$_2$In from magnetization measurements. The inset shows the exponent $n$ ($|\Delta S_T|\propto H^n$) \textit{vs.} $T$. (c) $\Delta S_T$ of Pr$_2$In from $MT$ measurements and heat capacity measurements. (d) $\Delta T_{ad}$ from heat capacity measurements. (e) (f) $\Delta S_T$ and $\Delta T_{ad}$ for light and heavy rare-earth-based \ce{R2In} \cite{Biswas.2022b,Biswas.2022,Guillou.2018,Zhang.2011,Zhang.2009,Zhang.2009b,Zhang.2009c}, \ce{RAl2} (Pr, Nd, Gd, Tb, Dy, Ho, Er) \cite{Liu.2022,Liu.2023,Gil.2016} in magnetic fields of 5 T. The shadows mark the range of 77 $\sim$ 20 K.}
    \label{Fig.1}
\end{figure*}

This part focuses on the magnetocaloric properties of \ce{Pr2In}. \textbf{\Cref{Fig.1}} (a) displays the magnetization ($M$) \textit{vs.} temperature ($T$) curves of Pr$_2$In in magnetic fields of 0.02, 1, 2, 5, 10 T. Two transitions are observed: one at \qty{56}{\kelvin} and the other at about \qty{35}{\kelvin}. The transition at about \qty{35}{\kelvin} was reported to be a possible spin reorientation transition \cite{Biswas.2020}. The transition at \qty{56}{\kelvin} was reported to be a first-order magnetic phase transition with an excellent $\Delta S_T$ of \qty{15}{\joule\per\kelvin\per\kilogram} in magnetic fields of \qty{2}{T} \cite{Biswas.2020}. \par

\textbf{\Cref{Fig.1}} (b) presents the $\Delta S_T$ of Pr$_2$In as a function of temperature in magnetic fields of 0.5, 1, 1.5, and \qty{2}{\tesla}. $\Delta S_T$ is calculated from MT measurements (shown in the inset in \textbf{\Cref{Fig.1}} (a)) with a magnetic field step of \qty{0.25}{\tesla}. This calculation is based on the Maxwell relation via the equation $\Delta S_T = \int_0^H  \mu_0 (\partial M / \partial T )_H \, dH $ \cite{Tishin.2003}. The $\Delta S_T$ calculated from MT measurements reaches \qty{17.5}{\joule\per\kelvin\per\kilogram} in magnetic fields of \qty{2}{\tesla} at 56.5 K, which is slightly higher than the value reported in Ref. \cite{Biswas.2020}. To confirm that the nature of the phase transition at about \qty{56}{\kelvin} is first-order, we calculated the exponent $n$ from the power law $\Delta S_T \propto H^n$ \cite{Law.2018} and plotted it as a function of temperature in the inset in \textbf{\Cref{Fig.1}} (b). The $n$ values in all fields overshoot 2, confirming the nature of the first-order phase transition. \par

$\Delta S_T$ can also be obtained from the S(T,H) curves constructed from the heat capacity data by equation $S(T, H) = \int_0^T \mu_0 (C_p(T, H)/T)\, dT$ \cite{Tishin.2003}.  After constructing the S(T,H) curves, $\Delta S_T$ is calculated by \cite{Pecharsky.1999}: 
\begin{equation}
    	\Delta S_T (T, H) = S(T,H) - S(T,0).
\end{equation}
The detailed procedure for calculating $\Delta S_T$ from heat capacity data is included in the supplementary \cite{Supplementary}. \textbf{\Cref{Fig.1}} (c) plots the $\Delta S_T$ obtained from heat capacity data in magnetic fields of 1, 2, 5, 10 T and $\Delta S_T$ from MT measurements in magnetic fields of 1 and 2 T. The $\Delta S_T$ from heat capacity measurements matches well with the $\Delta S_T$ from magnetization measurements, confirming the accuracy of the heat capacity measurements. In magnetic fields of 10 T, $\Delta S_T$ reaches a value of about \qty{25}{\joule\per\kelvin\per\kilogram}, and a plateau-like step emerges on the peak of the $\Delta S_T(T)$ curves, which is a character of first-order phase transitions \cite{Smith.2012}. \par

\textbf{\Cref{Fig.1}} (d) shows the $\Delta T_{ad}$ indirectly obtained from heat capacity measurements in magnetic fields of 1, 2, 5, 10 T. $\Delta T_{ad}$ is obtained from the constructed $S(T,H)$ curves via \cite{Pecharsky.1999}: 
\begin{equation}\label{dT}
    	\Delta T_{ad} (T=T(S,0),H) = T(S,H) - T(S,0),
\end{equation}
where $T(S,H)$ is the inverse function of $S(T,H)$. The detailed process for calculating $\Delta T_{ad}$ from the heat capacity data is included in the supplementary \cite{Supplementary}. In magnetic fields of 2 and 5 T, the $\Delta T_{ad}$ of Pr$_2$In reach 2 and 4.3 K, respectively.  \par

However, these two values are not as impressive as the remarkable $\Delta S_T$ in \ce{Pr2In}. \textbf{\Cref{Fig.1}} (e) and (f) compare $\Delta S_T$ and $\Delta T_{ad}$ of \ce{Pr2In} with other light and heavy rare-earth-based magnetocaloric materials in magnetic fields of 5 T. The $\Delta S_T$ of \ce{Pr2In} is not only significantly larger than that of \ce{Er2In}, but also larger than Pr$_{0.75}$Ce$_{0.25}$Al$_2$, which shows the largest $\Delta S_T$ among the light rare-earth-based Laves phase \ce{RAl2} series, and the heavy rare-earth-based Laves phase \ce{DyAl2}, known as a promising candidate for magnetocaloric hydrogen liquefaction \cite{Ribeiro.2021}. But \ce{Pr2In} has a much smaller $\Delta T_{ad}$ than \ce{DyAl2} despite that \ce{Pr2In} shows a larger $\Delta S_T$. The $\Delta T_{ad}$ of \ce{DyAl2} is about 1.5 times as large as that of \ce{Pr2In}. \par

\begin{figure*}[ht!]
    \centering
    \includegraphics[trim = {0 1cm 0 0},clip, width = \linewidth]{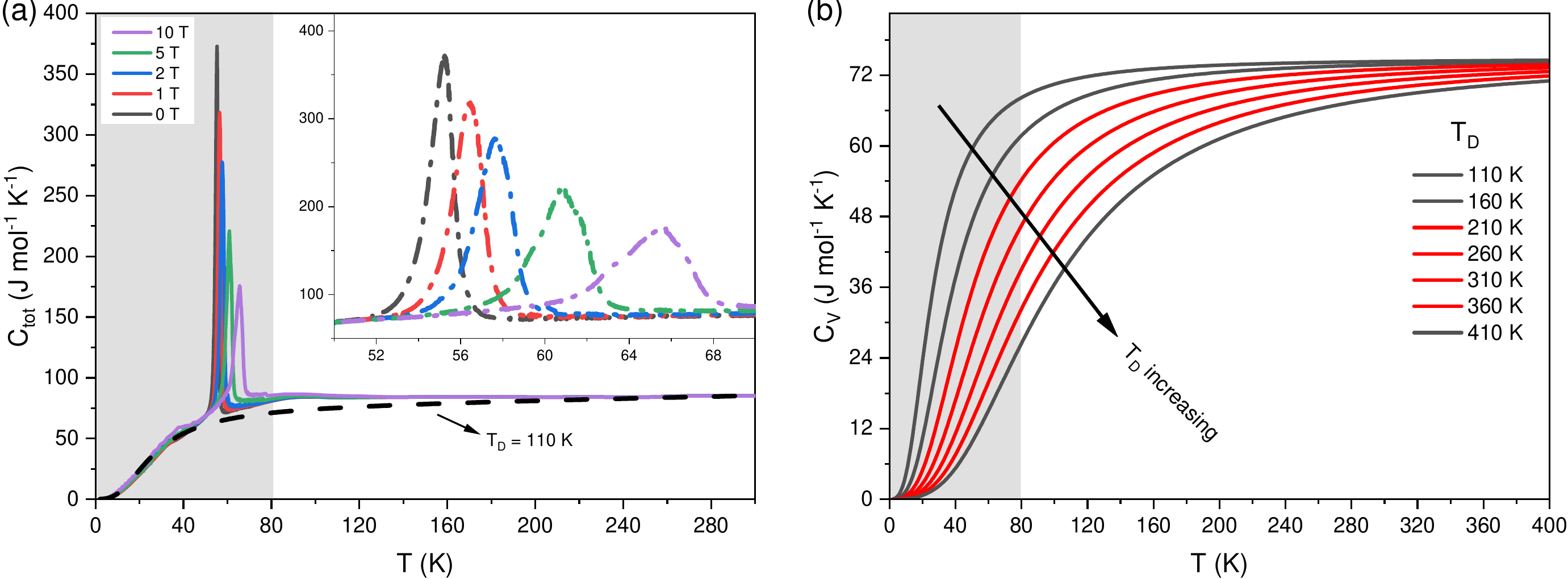}
    \caption{ (a) Heat capacity of Pr$_2$In as a function of temperature in magnetic fields of 0, 1, 2, 5, 10 T. (b) Volumetric lattice heat capacity from Debye model with $T_D$ varying from 110 to \qty{410}{\kelvin} with a step of \qty{50}{\kelvin}}
    \label{Fig.2}
\end{figure*}

Since $\Delta T_{ad}$ is indirectly obtained from the heat capacity measurement, a close look is given to the heat capacity data. \textbf{\Cref{Fig.2}}  (a) shows the total isobaric heat capacity $C_{tot}$ of \ce{Pr2In} in magnetic fields of 0, 1, 2, 5, and 10 T. One observation is that the peak of the heat capacity curves shifts with $H$, implying a first-order phase transition [36]. Another observation is that $C_{tot}$ of \ce{Pr2In} is almost constant even near \qty{80}{\kelvin}, indicating a low Debye temperature $T_D$. Due to the two magnetic phase transitions at low temperature, it is difficult to obtain $T_D$ from the linear relation $C_{tot}/T \, \propto \, \alpha T^2 + \gamma$ ($\alpha$ is the slope in which $T_D$ can be calculated, and $\gamma$ is the Sommerfeld coefficient) \cite{Kittel.2018}. In the literature, it is common to use the Debye model to fit heat capacity data to obtain $T_D$ \cite{Cwik.2021,Cwik.2019,Cwik.2022b}. This approach is based on the following equation:
\begin{equation}\label{cpfit}
    C_V + C_e = 9 N k_B \left ( \frac{T}{T_D} \right ) ^3 \int_0^{T_D/T} \frac{x^4 e^x}{(e^x - 1)^2}dx + \gamma T\, ,
\end{equation}
where $N$ is the number of atoms, $C_V$ is the volumetric lattice heat capacity, $C_e$ is the electronic heat capacity, $k_B$ is the Boltzmann constant, and $x=h\nu / k_B \, T$ with $\nu$ to be the frequency of the phonon. \par

In the present work, we obtained a Debye temperature of around \qty{110}{\kelvin} for \ce{Pr2In} using \Cref{cpfit}. This value is small, being outside the range of 200 $\sim$ \qty{400}{\kelvin} where $T_D$ of most alloys lie \cite{Kittel.2018}. A similar small value of about \qty{120}{\kelvin} was also reported for \ce{Yb2In}, an intermetallic compound that adopts the same crystal structure as \ce{Pr2In} \cite{Guillou.2020}.  \textbf{\Cref{Fig.2}} (b) plots the the volumetric lattice heat capacity $C_V$ for different Debye temperatures from 110 to \qty{410}{\kelvin} with a step of \qty{50}{\kelvin} using the Debye model. A significant difference between $C_V$ at cryogenic temperatures and near room temperature is revealed: $C_V$ for $T_D  \leq$ \qty{410}{\kelvin} at \qty{300}{\kelvin} are close, but at cryogenic temperatures such as 60 K, $C_V$ for $T_D <$\qty{210}{\kelvin} shows a significantly higher value. The difference between $C_V$ at cryogenic temperatures and near room temperature for different $T_D$ has led us to think about how $\Delta T_{ad}$ correlates with $T_D$. \par

Neglecting the electronic entropy as it is usually small compared to the magnetic entropy $S_m$ and the lattice entropy $S_l$ [45], the total entropy can be calculated by
\begin{equation}\label{totalS}
    S(T,H) = S_m + S_l 
\end{equation}
The magnetic entropy is given by \cite{Tishin.2003,Gottschall.2019}:
\begin{equation}\label{magneticS}
    S_m = N_M k_B \left [ \ln{\frac{\sinh{\left ( \frac{2J+1}{2J}y \right )}}{ \sinh{ \left ( \frac{1}{2J}y \right ) } }} - yB_J(y) \right ] \, ,
\end{equation}
with $N_M$ the number of “magnetic atoms”, $J$ the total angular momentum, $B_J (x)$ the Brillouin function, and 
\begin{equation}
    y = \frac{g_J J\mu_B \mu_0 H + \frac{3J}{J+1}k_B T_C B_J(y)}{k_BT} \, ,
\end{equation}
where $g_J$ is the Landé g-factor, $T_C$ the Curie temperature, $\mu_0$ the vacuum permeability.

The equation to calculate the lattice entropy $S_l$ is given by \cite{Balli.2017}:
  \begin{equation}\label{latticeS}
  \begin{split}
      S_l = & -3Nk_B \left[ \ln{\left ( 1 - \exp{\left ( -\frac{T_D}{T} \right )} \right) } \right] \\
      &+12Nk_B \left( \frac{T}{T_D}\right)^3 \int_0^{T_D/T} \frac{x^3}{\exp{(x)} - 1} dx \, .
  \end{split}
  \end{equation}
  \par
  
\Cref{dT,totalS,magneticS,latticeS} connect $\Delta T_{ad}$ with $T_D$. By varying $T_D$ and $T_C$, we can see how $\Delta T_{ad}$ changes. However, it should be emphasized that these equations only take $J$, $g_J$, $T$, $T_C$, and  $T_D$ as variables. In the present work, we only consider these parameters, ignoring the rest of the factors such as microstructures and stoichiometry that influence $\Delta T_{ad}$. In the present work, the values of $J$ and $g_J$ are taken as 4 and 4/5, respectively, which corresponds to \ce{Pr^{3+}}.\par

\begin{figure*}[ht!]
    \centering
    \includegraphics[trim={1cm 0cm 2.9cm 1cm}, clip, width = \linewidth]{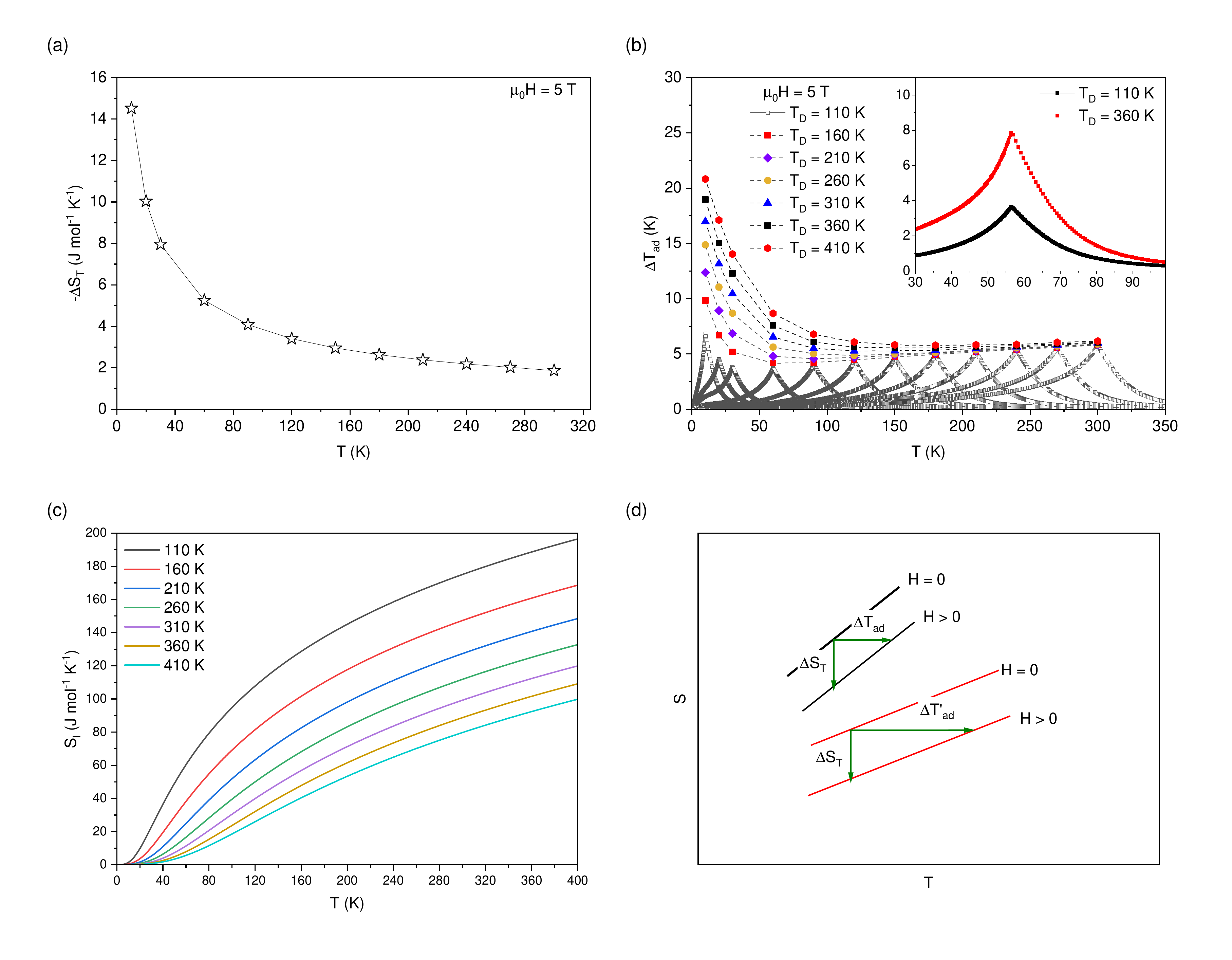}
    \caption{ (a) $\Delta S_T$ calculated from the mean-field approach. (b) $\Delta T_{ad}$ calculated from the mean-field theory with $T_C$ and $T_D$ varying. The inset compares the $\Delta T_{ad}$ for $T_D = 110$ K and 360 K with both $T_C = 56.5$ K. (c) Lattice entropies for different $T_D$. (d) Illustration of how the slope of the entropy curve influences $\Delta T_{ad}$. }
    \label{Fig.3}
\end{figure*}

The calculated $\Delta S_T$ and $\Delta T_{ad}$ using \Cref{dT,totalS,magneticS,latticeS} with $T_D$ and $T_C$ varying are displayed in \textbf{\Cref{Fig.3}} (a) and (b). \Cref{magneticS} implies that $\Delta S_T$ does not depend on $T_D$, $\Delta S_T$ should have the same value at the same $T_C$ regardless of how $T_D$ is varying. This is the reason why there is only one $\Delta S(T_C, T_D)$ curve in \textbf{\Cref{Fig.3}} (a). However, this is not the case for $\Delta T_{ad}$. In \textbf{\Cref{Fig.3}} (b), different $T_D$ leads to different $\Delta T_{ad}(T_C)$ curve. It can be observed that $T_D$ influences the turning point where the decreasing trend of the maximum $\Delta T_{ad}$ with respect to the decreasing $T_C$ turns to an increasing trend: for $T_D$= 110 K, the increasing trend is not observed until 30 K, while for $T_D$= 410 K, an increasing trend starts at 120 K. It can be concluded that material systems with higher $T_D$ tend to exhibit larger $\Delta T_{ad}$, particularly in the cryogenic temperature range. From the inset in \textbf{\Cref{Fig.3}} (b), the $\Delta T_{ad}$ of the material with a $T_D$= 360 K is more than twice as large as the material with a $T_D$ = 110 K, although both have the same maximum $\Delta S_T$ at 56.5 K. \par

It should be noted that there are no ideal material systems that only vary in $T_C$ and $T_D$ while keeping the remaining parameters constant. Furthermore, although the correlation between $\Delta T_{ad}$ and $T_D$ can be well described by the approach that combines the mean-field theory and the Debye model, further improvements are needed to include the factor of the nature of the phase transition order for a more profound interpretation. In particular, the nature and mechanism of the first-order phase transition of \ce{Pr2In} are not yet fully understood. Further theoretical and experimental investigations are required, such as the topological change of the Fermi surface of \ce{Pr2In} and its magnetic configurations. Moreover, $\Delta S_T$ and $\Delta T_{ad}$ are influenced by many factors, including extrinsic factors such as grain size and texture, and intrinsic factors such as crystalline electric field and stoichiometry\cite{Liu.2022,Lyubina.2017,Gutfleisch.2011,Gutfleisch.2016}. It should be also emphasized that the shifting of the transition temperature with respect to magnetic fields also influences $\Delta T_{ad}$ for first-order phase transitions \cite{sandeman_magnetocaloric_2012,porcari_reverse_2012,liu_giant_2012}. The relatively small $dT_C/dH$ (about 1 K/T for \ce{Pr2In}) also contributes to the absence of an excellent $\Delta T_{ad}$ in \ce{Pr2In}. Nevertheless, the mean-field approach presented in this work provides a way of understanding the absence of an outstanding $\Delta T_{ad}$ in \ce{Pr2In}. \par 

For a more generic interpretation on how $T_D$ influences $\Delta T_{ad}$, we consider the total entropy curve. \textbf{\Cref{Fig.3}} (c) shows the lattice entropy $S_l$ for different $T_D$. As observed, in cryogenic temperature range, the $S_l$ curve for smaller $T_D$ tends to exhibit a larger slope. Supposing that $S_m$ are all the same for all the $T_D$, it can be concluded that in cryogenic temperature range, the slope of the total entropy $S(T,H)$ is larger for smaller $T_D$, since 
\begin{equation}
    \frac{dS(T, H)}{dT} = \frac{dS_l (T)}{dT} + \frac{dS_m(T, H)}{dT}  .
\end{equation}
As illustrated in \textbf{\Cref{Fig.3}} (d), a steeper $S(H,T)$ results in a smaller $\Delta T_{ad}$, although both of them have the same $\Delta S_T$. In addition, based on the fact that it is a characteristic of first-order phase transition that the peak of the heat capacity shifts with magnetic fields, another explanation of how $T_D$ influences $\Delta T_{ad}$ for first-order phase transition is included in the supplementary \cite{Supplementary}.

\section{Conclusions}
In this study, the $\Delta T_{ad}$ of \ce{Pr2In} showing a first-order magnetic phase transition with an excellent $\Delta S_T$ is obtained indirectly from heat capacity data: 2 and 4.3 K in magnetic fields of 2 and \qty{5}{\tesla}, respectively. Motivated by the observation that the $\Delta T_{ad}$ of \ce{Pr2In} is not as significant as its $\Delta S_T$, research on \ce{Pr2In} continues to explain why an outstanding $\Delta T_{ad}$ in \ce{Pr2In} is absent. Inspired by the finding that \ce{Pr2In} shows a low $T_D$ of around \qty{110}{\kelvin}, the correlation between $\Delta T_{ad}$ and $T_D$ is studied. Combining the mean-field model with the Debye model, it is shown that $T_D$ has a substantial impact on $\Delta T_{ad}$: materials with a higher $T_D$ tend to show a larger $\Delta T_{ad}$, particularly at cryogenic temperatures. \par

Our work makes a connection between $T_D$, an important physical quantity that correlates the elastic properties with the thermodynamic properties (such as phonons, thermal expansion, thermal conductivity, specific heat, and lattice enthalpy), and $\Delta T_{ad}$, one of the most important magnetocaloric parameters \cite{li_computational_2012}. The important role of $T_D$ in achieving large $\Delta T_{ad}$ at cryogenic temperatures is demonstrated, which could guide the search or design of materials with both large $\Delta S_T$ and $\Delta T_{ad}$ by considering materials with high $T_D$. Furthermore, more research is required on the mechanism of the magnetocaloric effect in \ce{Pr2In}, since it is not yet fully understood. We should also explore ways to replace Indium as it is also a highly critical element. \\ \par

\section{Data availability statements}
The data that support the findings of this study are available upon reasonable request from the authors.

\section{Acknowledgement}
We gratefully acknowledge the supports from HLD (Dresden High Magnetic Field Laboratory), Deutsche Forschungsgemeinschaft (DFG, German Research Foundation) through the CRC/TRR 270 (Project ID 405553726 and ID 456263705), from European Research Council (ERC) under the European Union’s Horizon 2020 research and innovation program (Grant No. 743116, Cool Innov), and the Clean Hydrogen Partnership and its members within the framework of the project HyLICAL (Grant No. 101101461). We greatly appreciate the constructive discussions and useful experimental help from Marc Strassheim and Eduard Bykov from HLD, and Alex Aubert from TU Darmstadt. 

\bibliography{manuscript}% Produces the bibliography via BibTeX.

\end{document}